# ON QUANTUM INTEGRABLE SYSTEMS

V. Danilov, Oak Ridge National Laboratory, Oak Ridge, TN 37830
S. Nagaitsev, Fermi National Accelerator Laboratory, Batavia, Il 60510

*Abstract*

Many quantum integrable systems are obtained using an accelerator physics technique known as Ermakov (or normalize variables) transformation. This technique was used to create classical nonlinear integrable lattices for accelerators and nonlinear integrable plasma traps. Now, all classical results are carried over to a nonrelativistic quantum case.

## I. INTRODUCTION

Chaotic motion is not desirable in particle accelerators or plasma traps because it leads to particle losses and beam blow-ups. The preferred lattices in use are those that conserve simple invariants of motion and in most of them the motion is linear. The invariants of linear time-dependent oscillator were found by V. Ermakov [1] in 1880s and became known as Courant-Snyder invariants in accelerators [2]. Also, there exist a simple change of coordinates and time that results in a time-independent motion in special (so-called normalized) variables.

Surprisingly, this transformation (we will call it the Ermakov's transform below) allows one to create a large class of nonlinear integrable systems for accelerators [3] and plasma traps [4]. In this paper we extend it to quantum systems. We'll show that all classical time dependent integrable systems, found with the help of Ermakov transform, generate quantum integrable systems with a similar transformation of nonrelativistic quantum equations. The Schrödinger and Pauli equations become separable in some coordinate systems and can be solved in analytic functions.

## II. THE ERMAKOV'S TRANFORM

The equations of motion in accelerators (in the uncoupled case) are written as:

$$\begin{cases} x'' + K_x(s)x = 0 \\ y'' + K_y(s)y = 0 \end{cases} \quad (1)$$
$$K_{x,y}(s+C) = K_{x,y}(s)$$

where $K_x$ and $K_y$ are piecewise constant functions of $s$ (the time-equivalent longitudinal coordinate), and $C$ is the accelerator circumference and the longitudinal motion is negligible.

One can notice that these are two uncoupled Hill's equations. Such an equation was first solved by Ermakov [1] who obtained its invariant, which in accelerator physics is called the Courant-Snyder invariant [2]. This invariant can be understood by introducing the so-called normalized phase-space coordinates:

$$z_N = \frac{z}{\sqrt{\beta(s)}},$$
$$p_N = p\sqrt{\beta(s)} - \frac{\beta'(s)z}{2\sqrt{\beta(s)}}, \quad (2)$$

where $z$ stands for either $x$ or $y$, $p$ is similarly either $p_x$ or $p_y$, and $\beta(s)$ is either the horizontal or vertical beta-function (defined in, e.g. [5]). In these new normalized variables, the initial time-dependent Hamiltonian associated with Eqs. (1) becomes time-independent,

$$H = \frac{1}{2}\left(p_N^2 + z_N^2\right), \quad (3)$$

and thus leads to two invariants, the horizontal and vertical Hamiltonians (with new "time" $\mu$ with the following relation to the longitudinal coordinate $s$: $d\mu = ds/\beta(s)$ that is different for $x$ and $y$ planes in general). According to Eq. (3), in a linear lattice, all particles execute harmonic oscillations around the reference orbit with a frequency, known as the betatron tune, which is identical for all particles, regardless of their amplitude. Linear lattices have been considered attractive, in part because linear dynamics is easily understood. Now we use transform (2) for nonlinear systems. For any Hamiltonian of the form

$$H = \frac{p_x^2}{2} + \frac{p_y^2}{2} + K(s)\left(\frac{x^2}{2} + \frac{y^2}{2}\right) + V(x,y,s), \quad (4)$$

Transformation (2) produces new Hamiltonian

$$H_N = \frac{p_{xN}^2 + p_{yN}^2}{2} + \frac{x_{xN}^2 + y_{yN}^2}{2} \\ + \beta(\mu)V(x_N\sqrt{\beta(\mu)}, y_N\sqrt{\beta(\mu)}, s(\mu)) \quad (5)$$

where new "time" ($d\mu = ds/\beta(s)$) is the same for $x$ and $y$ planes). Now if we choose potential such that it is independent of $\mu$, we would obtain one invariant of motion – the new Hamiltonian (5). In addition, other invariants can be found to have completely integrable system (see [3, 4]).

## III. SCHRÖDINGER EQUATION

We rewrite the Schrödinger equation in some convenient form by redefining time and coordinates:

$$i\sqrt{2}\frac{\partial\psi}{\partial t} = -\nabla^2\psi + \frac{K(t)r^2}{2}\psi + U(\vec{r},t)\psi, \quad (6)$$

where time here is $t = \sqrt{2}T/\hbar$ (T is the real time), and coordinates $\vec{r} = \vec{R}\sqrt{2m}/\hbar$ ($\vec{R}$ stands for real coordinates), $K(t)$ is an arbitrary linear focusing coefficient and $U(\vec{r},t)$ is an arbitrary potential.

Now we introduce new "time" $\tau$, coordinates, and wave function by using functions $f(t)$, $\beta(t)$, and $g(t)$ (equations for them will be given below) with the following relation to conventional time and coordinates:

$$d\tau = \frac{dt}{\beta}, \quad \vec{r}_N = \frac{\vec{r}}{\sqrt{\beta(t)}}, \\ \psi_N = f(t)e^{ig(t)r_N^2}\psi \quad (7)$$

The equation (6) in new coordinates and time for new wavefunction becomes:

$$i\sqrt{2}\frac{\partial\psi_N}{\partial\tau} = -\nabla_N^2\psi_N + \frac{r_N^2}{2}\psi_N \\ + \beta(\tau)U(\vec{r}_N\sqrt{\beta(\tau)}, t(\tau))\psi_N \quad (8)$$

in case $\beta(t)$ obeys:

$$(\sqrt{\beta})'' + K(t)\sqrt{\beta} = \frac{1}{\beta^{3/2}}, \quad (9)$$

and

$$g = \frac{\beta'}{4\sqrt{2}}, \quad (10)$$

$$f = \beta^{-\frac{1}{4}}. \quad (11)$$

One can see a complete analogy with the Hamiltonian (5). In addition, Eq. (9) is exactly the envelope equation for linear motion in accelerators (see [5]), implying that and all classical results on integrability from references [3, 4] are applicable to Schrödinger equation. Specifically, all integrable time-dependent classical systems from those papers generate the time independent Schrödinger's equation in the new variables and can be solved analytically. In addition to transformation (7), there exist another transformation of simpler type (see 1st paper in Ref. [3], Eq. 25) that produces a similar conversion of Schrödinger equation to the same equation but with a different potential.

The simplest example of the conversion to a time-independent integrable system is given by $U(\vec{r},t) = A(\theta)/r^2$ in (6), where $A(\theta)$ an arbitrary function of a zenith angle in the spherical coordinates. Then, by solving Eq. (9) and transforming variables and the wave function using (7), (10), and (11), one obtains a time-independent equation (8) with the new potential (including quadratic term), $V(\vec{r}_N,\tau) = \frac{r_N^2}{2} + \frac{A(\theta)}{r_N^2}$ which is a classical integrable potential and separable in spherical coordinates. Note that the all results of a classic paper [6] are obtained in a single line (transformation (7)) and many more integrable systems can be found by this transform from classical results of papers [3, 4].

## IV. PAULI'S EQUATION

This equation is more involved than the previous one and but the transformation from time-dependent systems to time-independent ones is similar. The

Pauli equation (with the same redefinition of coordinates and time as in (6)) reads:

$$i\sqrt{2}\frac{\partial \varphi}{\partial t} = ((i\vec{\nabla} + d\vec{A})^2 + U(\vec{r},t))\varphi + \frac{e\hbar}{2m}\vec{\sigma}\cdot\vec{B}(\vec{r},t)\varphi \quad , (12)$$

where $\varphi$ is a two-component spinor wavefunction, $\vec{\sigma}$ are Pauli matrices, $U, \vec{A}$ are the electric and vector potentials, respectively, and $\vec{B}$ is the magnetic field, $d = e/\sqrt{2m}$. With the transformation (7) (with bispinor $\varphi_N$ instead of $\psi_N$) the Pauli equation is transformed to:

$$i\sqrt{2}\frac{\partial \varphi_N}{\partial \tau} = (i\vec{\nabla} + d\sqrt{\beta(\tau)}\vec{A}(\vec{r}_N\sqrt{\beta(\tau)},t(\tau)))^2\varphi_N$$
$$+ \frac{r_N^2}{2}\varphi_N + \beta(\tau)U(\vec{r}_N\sqrt{\beta(\tau)},t(\tau)))\varphi_N \quad (13)$$
$$+ (\frac{e\hbar\beta(\tau)}{2m}\vec{\sigma}\cdot\vec{B}(\vec{r}_N\sqrt{\beta(\tau)},t(\tau)) + 2A\sqrt{\beta}g\,x_N)\varphi_N$$

Equation (13) has a few new features with respect to the magnetic field. While the electric potential gets factor $\beta$ after the transform, the vector potential gains a factor $\sqrt{\beta}$ - they transform differently. It means, for example, that the electric potential $U \propto \frac{1}{r^2}$ is invariant under transformation (7), but for the vector potential the invariant function is $\vec{A} \propto \frac{\vec{a}}{r}$, where $\vec{a}$ is an arbitrary vector that doesn't change when all coordinates are multiplied by same arbitrary factor. For the magnetic field, though, the invariant dependence on coordinates is $1/r^2$ as it should be because $\vec{B} = \vec{\nabla}\times\vec{A}$. One can check that if term with vector and electric potentials is independent on time, the last two terms in (13) also don't depend on time, therefore we restore all the properties of Ermakov's transformation even for Pauli's equation – all the classical approaches on how to obtain time-independent integrable systems automatically work for this equation as well.

## V. CONCLUSIONS

In this paper we have described an extension of Ermakov-like transformation to the Schrödinger and Pauli equations. It is shown that these newly found transformations create a vast variety of time dependent quantum equations that can be solved in analytic functions, or, at least, can be reduced to time-independent ones.

## VI. ACKNOWLEDGMENTS